%% file: apssamp.tex
\documentclass[%
 reprint,
 amsmath,amssymb,
 aps,
prd
]{revtex4-2}

\usepackage{graphicx}
\usepackage{dcolumn}
\usepackage{bm}
\usepackage{booktabs}

\usepackage{caption}
\usepackage{subcaption}

\begin{document}

\preprint{APS/123-QED}

\title{Clouds as near-horizon backgrounds in the radio detection of air showers}


\author{Enrique Huesca Santiago}
\author{Giulia Cantarini}%
\author{Anna Nelles}
\affiliation{%
 Deutsches Elektronen-Synchrotron DESY, Platanenallee 6, 15738 Zeuthen, Germany\\
}%
\affiliation{%
 ECAP, Friedrich-Alexander-Universität Erlangen-Nürnberg, Nikolaus-Fiebiger-Str.~2, 91058 Erlangen, Germany\\
}

\date{\today}

\begin{abstract}
Stand-alone radio detection of air showers from cosmic rays or neutrinos relies on effective and efficient suppression of human-made noise sources. We propose here that reflections of human-made transmitters on clouds or other atmospheric phenomena, can cause confusion backgrounds that have not been studied previously. Given the absence of prior modelling, we have developed a simplified model for the reflection strength of clouds in the MHz regime. The results indicate that such reflections are, in principle, detectable, in particular in extremely horizontal geometries with typical experimental set-ups. 
\end{abstract}

\maketitle

\section{\label{sec:1}Introduction}
\input{sections/01_introduction}

\section{\label{sec:RFI_evidence}Evidence of unexplained RFI}
\input{sections/02_experimental_evidence}

\section{\label{sec:level3}Common model for horizon RFI}
\input{sections/03_RFI_model}

\section{\label{sec:level4}The cloud model}
\input{sections/04_cloud_model}

\section{\label{sec:level5}Conclusions}
\input{sections/05_conclusions}

\section{Acknowledgements}
We thank our colleagues in the radio community, in particular Sjoerd Bouma, for insightful discussions, and John Kelley and Dima Kostunin, for digging into self-triggered data and discussing trigger details with us. Even after presenting our idea at \cite{HuescaSantiago:2026hf} no concerns were raised that killed it, which motivated us to publish this article. We also thank Karen Terveer helping us with her graphical skills to generate Figure 6. 

The authors acknowledge funding from the European Union, under grant Pro-RNO-G (No 101115122). Views and opinions expressed are however those of the author(s) only and do not necessarily reflect those of the European Union.

\appendix

\section{\label{app:radio_propagation_in_atmosphere} Atmospheric radio refraction}

\input{sections/A_radio_propagation_atmosphere}


\bibliography{apssamp}

\end{document}

%% file: sections/01_introduction.tex
While radio detection of air showers has matured as a technique, the problem of a highly efficient and pure self-trigger has not been universally solved. This is mostly due to the abundance and variety of radio background signals (RFI) as a function of location and time that cause false-positive triggers, often at rates that saturate data-acquisition systems. While this fact has caused radio detection of air showers to be almost abandoned in times of analogue-only electronics, the advent of digital electronics and the feasibility of more complex triggers have enabled vetoing backgrounds on-the-fly \cite{Huege:2016veh}. 
Still, the \emph{horizon}, here referring to a couple of degrees above the physical horizon, is often excluded in triggers or analysis, simply because the background is overwhelming. If looking for cosmic ray induced air showers, this is certainly a viable path forward, given that the effective area lost is minor. However, for experiments looking for the radio emission of neutrinos, by detecting showers stemming from Earth-skimming (tau-)neutrinos \cite{MammenAbraham:2022xoc}, this exclusion is no longer feasible.

Radio detection of Earth-skimming neutrinos relies on the fact that detectors are constructed with superior angular resolution in mind that will be able to tell the difference between a background from or just above the horizon and the true signal from below the horizon. Many of the proposed instruments also exploit the topography of the Earth to provide a good view of \emph{below-horizon} Earth \cite{GRAND:2018iaj,Southall:2022yil,GRAND:2025rps}. One should remember that a location on top of a mountain changes where the horizon is, and we have tried to be very careful in this article when converting coordinate systems to one where $0^{\circ}$ is horizontal.  

When reviewing past experiments and their efforts towards self-triggering, we have noticed that many of them report unexpected signals from close to, but not compatible with, the horizon. Many experiments also report that these backgrounds are time-variable, impulsive, and often remain in the data after known terrestrial sources, airplanes, and thunderstorms have been removed.   

Here, we want to propose that clouds reflecting terrestrial emissions in skimming geometries may be the cause of backgrounds appearing as above-horizon sources. In this article, we infer the radar cross-section of clouds at MHz frequencies using simplified droplet models. Together with basic geometric considerations, we find that the hypothesis is not so easily rejected and in fact consistent with general assumptions about experimental sensitivity and cloud properties. We encourage experiments running a self-trigger to investigate whether they can experimentally test the hypothesis to understand whether clouds (or related atmospheric phenomena) can cause terrestrial backgrounds to be confused as not stemming from the horizon, which will provide further challenges to high-purity and high-efficiency self-triggers, in particular for those experiments located above the clouds.

%% file: sections/02_experimental_evidence.tex
We collect here the published evidence for above-horizon RFI and attempt to compare the observations. Clearly, all experiments use different triggers, and not all trigger details are publicly available. Thus, we were careful not to over-interpret the evidence at hand, but rather to use it as motivation for our study.

\subsection{AERA}
The Auger Engineering Radio Array (AERA) was the first medium-scale installation of MHz antennas in the Pierre Auger Observatory in Argentina \cite{PierreAuger:2025hoe}. It had several sizes, starting with an array of 24 antennas with a spacing of 100~m. 
Throughout its lifetime, AERA routinely operated triggered from the particle detectors of the Pierre Auger Observatory in Argentina \cite{PierreAuger:2023lkx}. However, in its first years of operation, a fraction of the stations were running a self-trigger as reported for instance in \cite{Kelley:2012agy}. With the self-trigger that used a dual-threshold crossing metric, a clear excess from the horizon in certain azimuthal directions was observed, as one would expect from terrestrial RFI sources. However, there is also a certain haze of events seen that peak at about $10^{\circ}$ above the horizon. At the time, this set of events was attributed to coarse reconstruction methods and a detector in commissioning \cite{Fliescher2011}. Thereby known RFI sources at the horizon, like power lines, villages, or oil drilling fields, would mis-reconstruct to above the horizon. 

\begin{figure}
    \centering
    \includegraphics[width=1\linewidth]{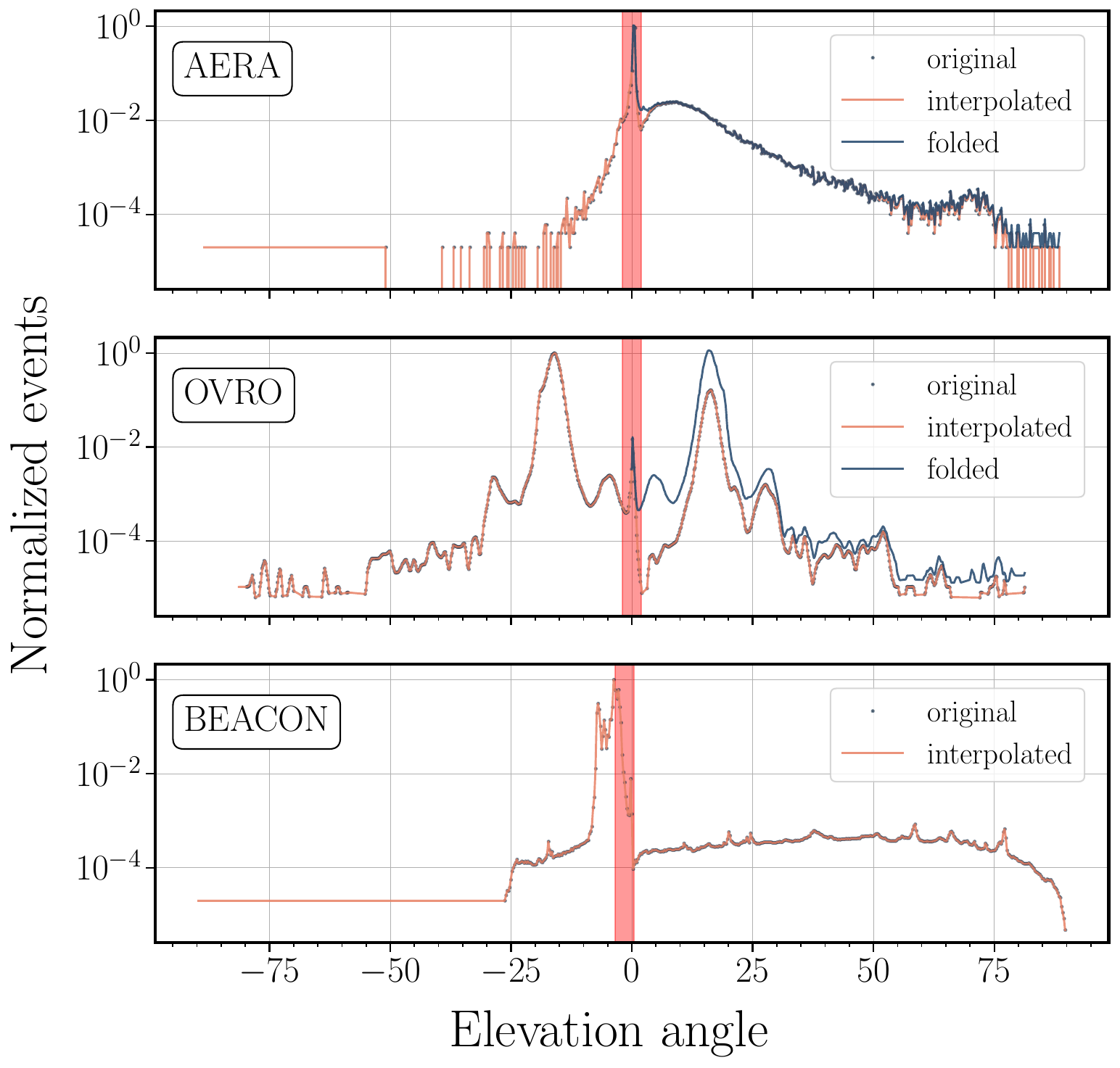}
    \caption{Distributions of arrival directions reported in self-triggered data from AERA, OVRO, and BEACON, using plane-wave approximations. The red band ($\pm 2^{\circ}$) marks the local horizon.}
    \label{fig:overview_experiments}
\end{figure}

As the self-trigger of AERA, in particular the central event building trigger, was further developed and ultimately for the Radio Detector of the Pierre Auger Observatory, the radio antennas were slaved to the particle detectors to limit the data rate \cite{Hoerandel:2025Km}, these backgrounds were likely suppressed somewhere in the chain, at least no further investigation was published. 
We extracted the angular distribution from \cite{Kelley:2012agy}, summing over all azimuth angles and depict these in Fig.~\ref{fig:overview_experiments}.
 
\subsection{OVRO-LWA}
The Long Wavelength Array at the Owen's Valley Radio Observatory (OVRO-LWA) is a versatile radio telescope in California \cite{Eastwood:2017lex}. Also the OVRO-LWA underwent several upgrades and extensions, but we focus here again on early data with 256 dual-polarization antennas over an area of roughly 1~km$^2$.  In one of its modes, OVRO-LWA successfully ran a powerful self-trigger, where the full data-stream of the antennas was combined at a central location \cite{Monroe:2019zkp}. Thereby, real-time suppression of RFI was possible and similarly the study of its parameters. The data published in \cite{Monroe:2019zkp} shows an excess at about $15^{\circ}$ above the horizon using a simple but robust plane-wave fit. Individual inspection of the events yielded that the time-differences were in disagreement with events straight from the horizon. Using sophisticated online tools the data was further reduced and cosmic ray detected. 

The authors note that the period of data taking was a rainy and cloudy period in California, which is known to suppress backgrounds from faulty powerlines. Since the OVRO-LWA is a shared instrument, the cosmic ray observations did not run in parallel to other observations, so only data from a limited period of less than two days was published.
Most of the observed above-horizon excess comes from the direction of nearby town Bishop, with airplanes and other known sources removed. The excess would correspond to a reflection at a height that corresponds to the reported cloud height on that day \cite{Kanitz2020}. The data published in the paper as histogram was added to Fig.~\ref{fig:overview_experiments}.

\subsection{BEACON}

A prototype of the Beamforming Elevated Array for COsmic Neutrinos (BEACON) was operated in the White Mountains of California \cite{Southall:2022yil}. Here, 4 dual-polarized antennas were installed at 1.5~km above the valley floor and operated in a phased-array mode. In \cite{Southall:2022yil} 112 days of raw data are evaluated and cosmic ray candidates were identified. 

For an experiment high above the ground, the geometry is notably different and care has to be taken to compare the findings. However, it also seems a similar excess was found that does not necessarily match with expectations form the horizon. The data as reported in \cite{Southall:2022yil} was added to Fig.~\ref{fig:overview_experiments}.

\subsection{GRANDProto300}

A prototype of the GRAND experiment \cite{GRAND:2018iaj}, GRANDProto300, is currently under construction in XaoDushan in the Gansu Province of China  \cite{Ma:2025lV}. This proto-type will consist of 300 so called HorizonAntennas, covering again the low MHz frequencies. 

In Figure 4 of \cite{Ma:2025lV} the reconstructed arrival directions of a subset of antennas are shown. It was noted in the corresponding presentation that also here an excess was seen that was not easily explainable. It changed with time and corresponds to the excess at $15^{\circ}$ above the horizon. The presentation even speculated about \emph{clouds}. Given that the data shown at the conference was preliminary and only covered 2 days, we did not attempt to extract the data from the figure, but note that the presentation of this result triggered our renewed interest in the subject. 

\subsection{Other radio detectors}
Discussion with colleagues from \textbf{Tunka-Rex} \cite{Tunka-Rex:2019msu} revealed that similar excesses above the horizon were seen during commissioning, but never published since these were not the target of the experiment. Their origin was never conclusively attributed to anything but suppressed in trigger and analysis. 

\textbf{LOFAR} routinely detects cosmic rays, however, never continuously ran a self-trigger \cite{Schellart:2013bba}, which would be needed to see such a background. Furthermore, almost all radio astronomical observations with LOFAR exclude sources too close to the horizon as it is known that RFI is not always contained to $0^{\circ}$, which explains that such a phenomenon may be known anecdotally, but was never consistently studied in interferometric or beamformed observations. 

For completeness, we should mentioned that the background situation at \textbf{LOPES} \cite{LOPES:2021ipp} and other very early experiments was so challenging that their data of rare backgrounds will be of limited usage without detailed reanalysis. 

The obvious candidate to detect such a phenomenon would be the \textbf{ANITA} payload that flew over Antarctica to search for ultra-high energy neutrinos \cite{ANITA:2019wyx}. However, the ANITA trigger was complex and the geometry of the problem continuously changing, given that the balloon moves. So we have not attempted to extract data to a common framework, also because the ANITA collaboration does not publish raw event distributions, but hope that our work triggers the interest of someone in the collaboration. It might also be useful for the successor of ANITA, \textbf{PUEO} \cite{PUEO:2020bnn}, to investigate the influence of clouds on RFI sources from their recent flight.  

All in-ice radio neutrino arrays also use a self-trigger, like \textbf{ARA} \cite{ARA:2022rwq}, \textbf{ARIANNA} \cite{Anker:2019rzo}, and \textbf{RNO-G} \cite{RNO-G:2024esr}. However, their usefulness for this study is limited. Due to their geometry, a signal has to refract into the ice before detection, which makes the distinction between from the horizon or just above extremely challenging, when ice properties are not perfectly known. Further, one of the most dominant backgrounds in in-ice detectors seems to be related to high-winds \cite{Aguilar:2023tba}, which in-turn probably also relates to clouds, making it hard to disentangle a potential correlation.

%% file: sections/03_RFI_model.tex
As described in Sec.~\ref{sec:RFI_evidence} we have curious evidence that there is a unexplained excess amount of events near the horizon, from which we can deduce a few properties to develop a common model of the origin. 

\subsection{Known properties of the radio excess }

First, we know that the background composed of radio signals on the 30 - 80(+) MHz band, which is the typical band of in-air astroparticle radio detectors. This signal is present in the results after triggering on impulsive characteristics, even surviving data cuts, so it has to be impulsive, compact in time, as opposed to a $O(\text{s})$ continuous wave signal of radio communications, FM radio, or TV transmission. 

Second, the excess is near-horizon, but clearly not at the horizon itself. The horizon we need to consider here is the radio horizon and not the optical one, so any region $\pm 2-5^\circ$ over the optical horizon is still compatible with the radio horizon, given the different properties of radio propagation in the atmosphere (c.f.~App.~\ref{app:radio_propagation_in_atmosphere}) and angular resolution of the experiments. By near-horizon, we refer here to a range of low elevation angles between $10^\circ$ and $20^\circ$ over the local experiment horizon.

This is the approximate region where we can see in Fig.~\ref{fig:overview_experiments} that the different experiments report their excess events. It should be noted that AERA and OVRO's negative elevation angles are simply a product of the degeneracy existent in a plane-wave reconstruction algorithm when all the antennas a located on the same plane. BEACON had a non-planar distribution of antennas and observed from a mountain, and therefore does not have zenith ambiguity. However, due to the height, the visual horizon appears at $-1.5^{\circ}$~\cite{Southall:2022yil} and therefore the data cannot be compared directly. 

The distributions suggest that the excess originates from a spatially incoherent source. The estimates of $\theta$ and $\phi$ come from well-understood position reconstruction algorithms. Yet, the near-horizon excess does not present as a definite point in a spatial distribution, unlike most triggered events. In, e.g., AREA, reconstructed sky maps place most events at the location of sources of radio frequency interference (RFI), such as a power line or a transformer substation. 

Third, the near-horizon excess is clearly sub-dominant over actual horizon triggers; as shown for AERA, the additional triggers are less than $10^{-2}$. The small amount of triggers hints at something of sporadic nature, an intermittent effect that produces only a small amount of triggers and is a secondary, if not tertiary background during commissioning or in a systematic study. Any persistent radio sources would have been identified. 

GRAND, at least, has mentioned the possibility that these excess events might be caused by air traffic \cite{Ma:2025lV}. 
With the public data, it remains unclear to us whether this is a viable explanation for GRAND. We do note, however, that the OVRO-LWA and BEACON data has been specifically cleaned of these backgrounds and AERA is in a location with only very few flights \cite{PierreAuger:2015aqe}.

To recap, local background situations are versatile and changing, but when comparing the distributions from multiple experiments qualitative evidence emerges that in addition to local peculiarities a more global background remains. 
In spite of being a sub-dominant excess, we argue that this effect requires further investigation, particularly in the field of transient astrophysics. Typical background suppression for efficient triggers has to achieve something like $10^{-4}$ in order to match expected rates of cosmic rays and down to $10^{-9}$ for neutrinos, which are much smaller than the backgrounds observed here. 

Unfortunately, while the qualitative evidence is large, these conclusions have to be kept at a global level. Further details require precise experiment-level knowledge and access to the waveform-level data (to investigate the frequency content, i.e.). It is not feasible for people from the outside of any of these experiments to study the intricate details in a precise, quantitative fashion. Currently no experiment publishes the fully raw data before cuts and only few the related distributions. 

\subsection{Cloud reflection hypothesis}

The properties of the excess presented in the section above, along with anecdotal evidence of the correlation of unusual event with weather conditions, led to the hypothesis that these sporadic, near-horizon events are caused by atmospheric radio interference. Most common types of atmospheric radio interference, like ducting, have already been explored as a possibility, but were discarded \cite{Kanitz2020}; see App.~\ref{app:radio_propagation_in_atmosphere} for a brief estimate of this effect. 

What we propose here is that these sporadic events in the low MHz band are actually nearby radio reflections from clouds. A radio reflection would mean that there is no need to have a source in the observed/reconstructed direction, but rather just a cloud. The actual source of radio (intentionally transmitting or not) might lie much further away, at a distance or position typically considered incompatible with a possible source; e.g., behind a mountain or beyond the optical horizon. 
The cloud reflection hypothesis would also explain the event frequency, as it would not be a constant phenomena but dependent on (type of) cloud coverage and geometry; and also explains the spatially incoherence of the signal, is not a fixed source of reflection. 

\begin{figure}
    \centering
    \includegraphics[width=1\linewidth]{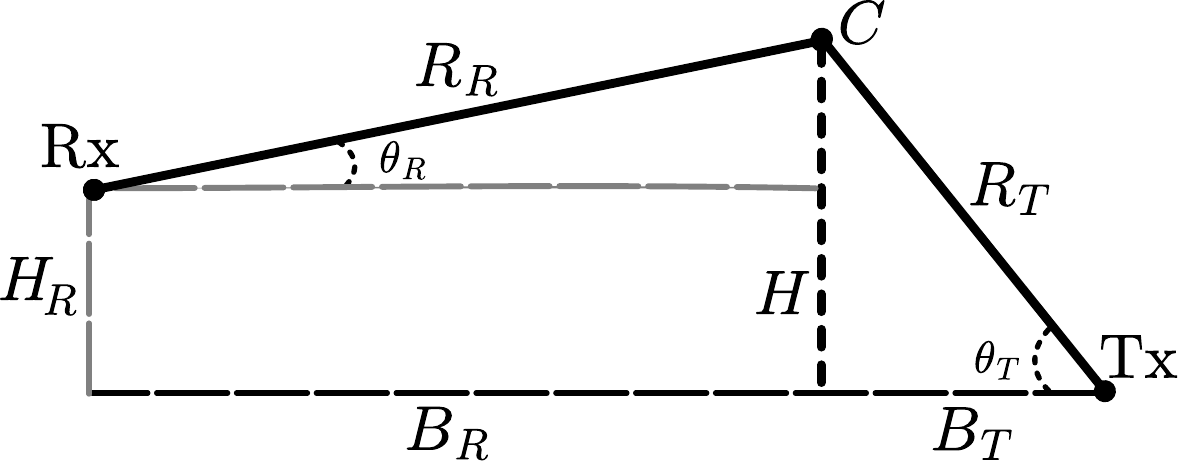}
    \caption{General geometry of the radio reflection. The radio wave travels from a transmitter ($Tx$, right) to the reflecting centre $C$ (at an altitude H over the ground), where its redirected to the receiver antenna ($Rx$, left), which may be at an altitude difference $H_R$ from the local ground. $Tx$ and $Rx$ have their respective viewing angles to the reflector, $\theta_T$ and $\theta_R$. $R = R_T + R_R$ is the total travel distance of the radio wave. $B = B_T + B_R$ is the total baseline or ground distance between source and receiver. }
    \label{fig:geometry}
\end{figure}

In order to explain the near-horizon geometry with cloud reflections, we first investigate if the near-horizon geometry is compatible with a typical cloud altitude. As a tropospheric effect, clouds are only expected to exist approximately in the first 15 km of the atmosphere, with their typical height  dependent on latitude \cite{NOAA}.

In Fig.~\ref{fig:geometry}, we present a simplified point-like reflection geometry for this scenario and we can determine that:

\begin{equation}
    H = B_R \tan(\theta_R) + H_R.
\end{equation}

\begin{figure}
    \centering
    \includegraphics[width=1\linewidth]{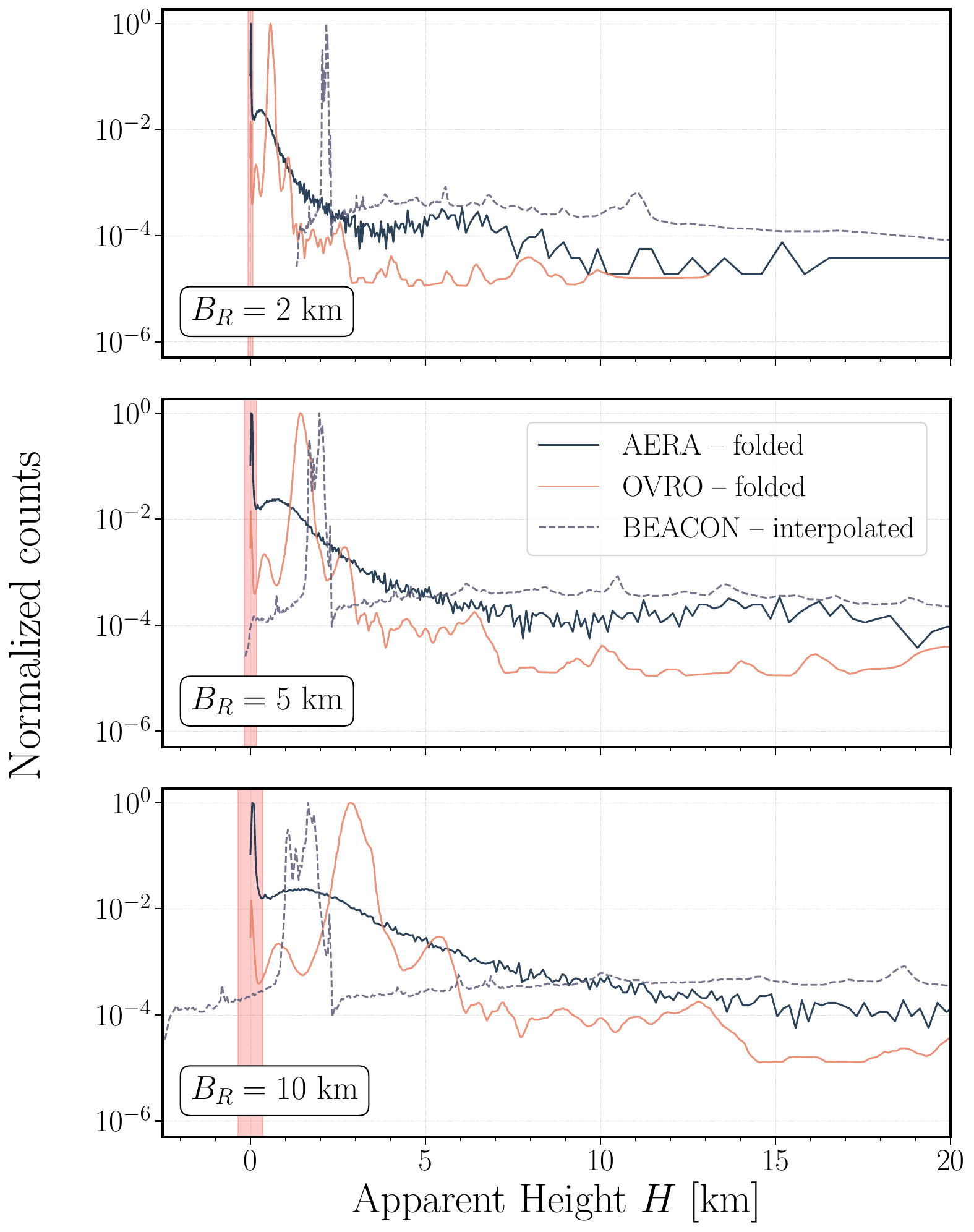}
    \caption{Measured arrival directions (same as in Fig.~\ref{fig:overview_experiments}) as function of apparent height using different assumption of distance $B_R$. The red band indicates as before the horizon at $\pm 2^{\circ}$.}
    \label{fig:height_distribution}
\end{figure}

While AREA and OVRO are taken at relative height $H_R = 0$, as they are located at the valley floor; BEACON is taken at  $H_R =$ 2.3 km over ``local floor" ($3.8 - 1.5$ km a.s.l. for the mountaintop and the valley floor, respectively). By taking a guess on $B_R$, we can convert the observed angles to an estimated reflection height $H$, as done for Fig.~\ref{fig:height_distribution}. 

We can notice two important, high-level details. First, the excess events are compatible with tropospheric heights for reasonably long baselines, around $< 100$~km. However, the hypothesis of a common model in the subdominant excess peaks in the different experiments is more supported by low-altitude ($<5$~km) and nearby ($\sim 10$~km) reflections. 

\begin{figure}
    \centering
    \includegraphics[width=1\linewidth]{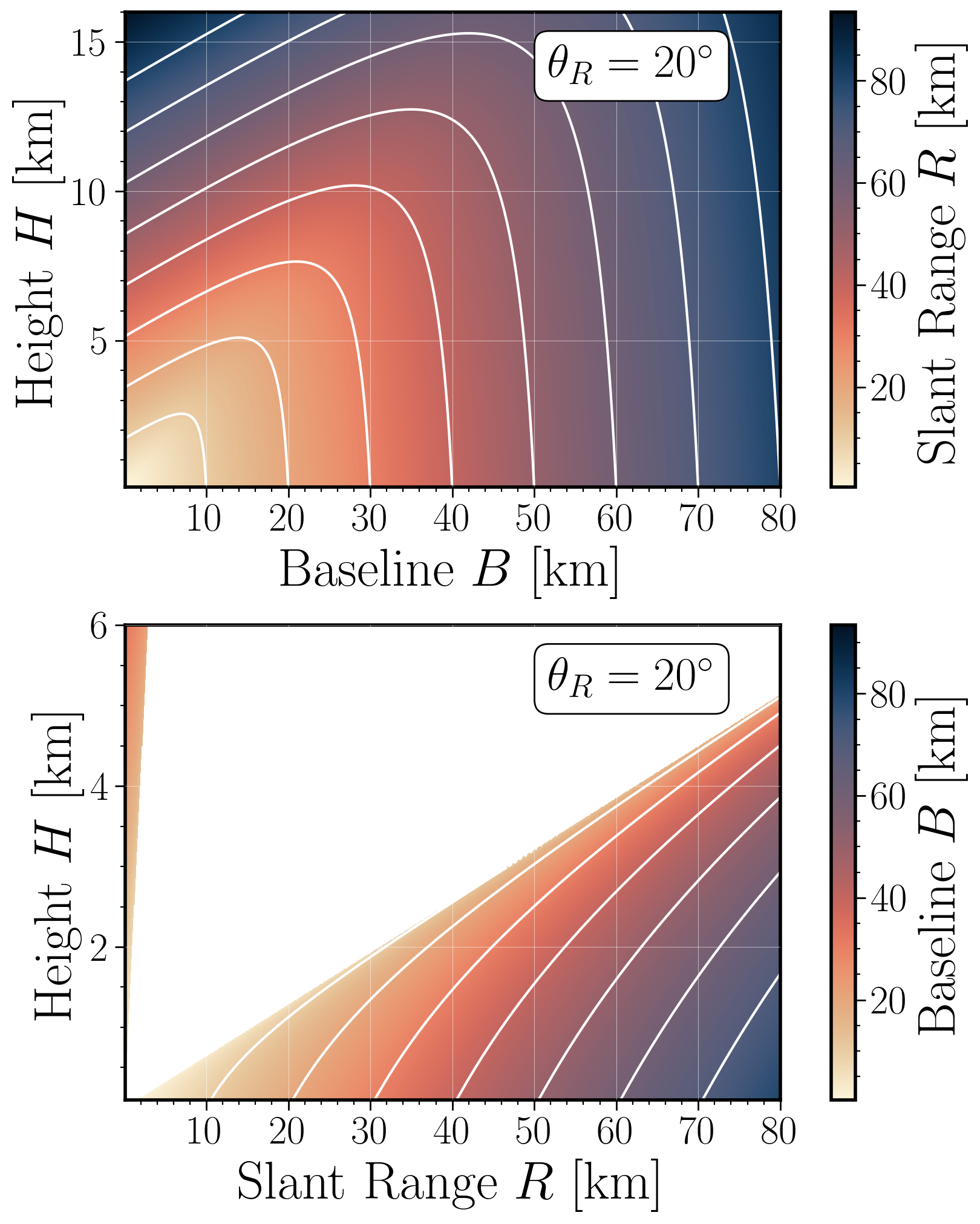}
    \caption{Available phase space for $B, H$ and $R$ (see Fig.~\ref{fig:geometry}) for a given observation angle $\theta_R$. The white lines are the lines of constant $B$ (hence why they converge to the x axis, $B=R$ when $H=0$).}
    \label{fig:range_phase_space}
\end{figure}

While the heights and the baseline are free parameters, they (together with $\theta_R$) define the total distance $R$ that the radio waves travel in order to make the reflection. By further inspecting Fig.~\ref{fig:geometry}, we can derive the following relationships:

\begin{equation}
\label{eq:R_vs_B_H}
\begin{split}
    R &= \sqrt{(B - B_R)^2 + H^2} + R_R; \\
R(B, H, H_R, \theta_R)  &= 
\sqrt{\left(B - \frac{H - H_R}{\tan(\theta_R)} \right)^2 + H^2 } 
+ \frac{H - H_R}{\sin(\theta_R)}. 
\end{split}
\end{equation}

\begin{equation}
\label{eq:B_vs_R_H}
\begin{split}
    B &= B_R + B_T = B_R + \sqrt{(R - R_R)^2 - H^2};  \\
B(R, H, H_R, \theta_R)  &= 
\sqrt{\left(R - \frac{H - H_R}{\sin(\theta_R)} \right)^2 - H^2} 
+ \frac{H - H_R}{\tan(\theta_R)} 
\end{split}
\end{equation}

These relations work even for $H < H_R$, as $H - H_R$ is negative but so is the sign of $\theta_R$, keeping the fraction terms  positive.  

For a receiver at floor level, $H_R = 0$, Eqs.~\ref{eq:R_vs_B_H} and \ref{eq:B_vs_R_H} reduce to:
\begin{equation}
\begin{split}
R(B, H, \theta_R)  &= \frac{H}{\sin(\theta_R)} \\
&\left(1 + \sqrt{1 - \frac{B}{H}\sin(2\theta_R) + \left(\frac{B}{H}\sin(\theta_R)\right)^2}
  \right) \\
B(R, H, \theta_R)  &= \frac{H}{\tan(\theta_R)} \\
&\left(1 + \sqrt{1 - \frac{R}{H}\frac{2}{\cos(\theta_R)} + \left(\frac{R}{H}\tan(\theta_R)\right)^2}
  \right)
\end{split}
\end{equation}

The available phase-space of solutions is presented in Fig.~\ref{fig:range_phase_space}. Baseline values as a function of $R$ and $H$ for a fixed $\theta_R)$. As Eq.~\ref{eq:B_vs_R_H} has a square root form, the white area corresponds to the unphysical negative radicand values. The left region, while not negative, implies that $B > R$, which is incompatible with our geometry. Only the right, coloured region is physically possible.   The values of $H$ are confined to $\simeq 15$ km, as we are assuming a reflection within the troposphere. This means that constraining the possible values $B, H$, requires an estimate on the maximum $R$ possible. This can be done estimating the energy budget of the radio reflection.

\subsection{Power budget and radar cross section}

Any radio reflection, is, by definition, equivalent to a radar system; radio systems that look for reflections are sometimes called passive or involuntary radars. The power balance along the radar system is calculated using the radar range equation:

\begin{equation}
\label{eq:radar_range}
    P_R = 
    \frac{\lambda^2}{(4\pi)^3 }
    \frac{P_TG_T G_R \sigma_{RCS}}{R_T^2R_R^2} 
    =
    \frac{c^2}{(4\pi)^3 f^2 }
    \frac{P_{\text{EIRP}} G_R \sigma_{RCS}}{R_T^2R_R^2}.
\end{equation}

The received power $P_R$ is defined at the antenna terminals (this is, before the RX amplification system). It is a function of the transmitter's effective isotropic radiated power (EIRP), the receiver's antenna gain $G_R$, the wavelength $\lambda$ or frequency $f$, the distances $R_T, R_R$ and the radar cross section $\sigma_{RCS}$ (more on that below). $P_{\text{EIRP}}$ is the combination of transmitter nominal power, $P_T$ and its total gain $G_T$. 

An alternative version can be written by taking the $10\log_{10}$ of each side,

\begin{equation}
\begin{aligned}
10\log_{10}(P_r) =
    \;& 
    10\log_{10}(P_{\text{EIRP}})
    + 10\log_{10}(G_R) 
    \\ &
    - 20\log_{10}(f)
    + 10\log_{10}(\sigma) 
    \\ &
    - 20\log_{10}(R_R)
    - 20\log_{10}(R_R) 
    \\ &
    - 10\log_{10}\left(\frac{\lambda^2}{(4\pi)^3 }\right),
    \\ 
P_R~[\mathrm{dBm}] =
    \;&
    P_{\text{EIRP}}~[\mathrm{dBm}]
    + G_R~[\mathrm{dBi}]
    \\ &
    - 20\log_{10}(f~\mathrm{[MHz]})
    + \sigma~\mathrm{[dBsm]}
    \\ &    
    - L_d(R_T~\mathrm{[km]}) - L_d(R_R~\mathrm{[km]})
    -103.44,
    \\ 
\end{aligned}
\end{equation}

This is the power-level dB radar range equation. Almost every term has been converted into its dB form, with $L_d(R) = 20\log_{10}(R)$ being distance loss terms  due to the simple inverse square law ($1/R^2$) (in dB). This formula is the easiest way to quickly estimate the \textit{power balance or budget} of the reflection. We have neglected other factors, such as long-distance air attenuation or cable losses. Still, this equation still has too many unknown variables to be useful to derive a general term for the maximum radar range. What we can do instead is estimate the range of values applicable to astroparticle radio telescopes, making assumptions about typical receiver values. 

We assume an operating frequency between 30 and 100 MHz, which makes the frequency loss term to be $[-30, -40]$ dB, approximately. The received signal power $P_R$ at a radio array cannot be detected with a standard threshold trigger if it is lower than the noise floor. That can be (optimistically) as low as $-100$~dBm, or 0.1 pW (picowatt), equivalent to a $V_R = 1 ~\mu \mathrm{W}$ for a Z = 50 $\Omega$ system, (before RX chain amplification). For simplicity, we take an effective value range of $P_R = -90$, to account for non-optimistic cases plus the margin for a threshold trigger $P_{R,\mathrm{trigger}} = P_R + \mathrm{SNR}_{\mathrm{trigger}}$. The events under investigation were strong enough to appear during the commissioning phase of all experiments, where triggering schemes were likely not optimized.  We can also take a typical $G_R = [10, 20]$ dB for the combined gain on a 100 unit antenna array under incoherent sum ($10$ dB) or coherent ($20$ dB, perfect beamforming). All of these are typical numbers for a highly-directional single-antenna gain.

Rearranging and substituting the equation above,

\begin{equation}
\label{eq:RCS_dB}
\begin{aligned}
- \sigma~\mathrm{[dBsm]} &=
    P_{\text{EIRP}}~[\mathrm{dBm}]
    \\ &
    - L_d(R_T~\mathrm{[km]}) - L_d(R_R~\mathrm{[km]})
   \\ &
    - [23.44, 43.44]
\end{aligned}
\end{equation}

The $\sigma_{RCS}$ cross section is kept with the minus sign as a convention here since the expectation is to detect a negative value of dBsm (decibel values over 1 square meter), corresponding to radar cross section values smaller than 1 square meter. The radar cross section (RCS) is a term introduced here as part of Eq.~\ref{eq:radar_range}, but not explained. The official definition is:
\begin{equation}
\label{eq:RCS_definition}
    \sigma
    = \lim_{R \to \infty} 4\pi R^2 \frac{S_s}{S_i},
    = \lim_{R \to \infty} 4\pi R^2 \frac{|E_s|^2}{|E_i|^2},   
\end{equation}
as the far-field ratio of the scattered power or energy density ($S_s$) to the incident power ($S_i$). In other words, it represents the efficiency of the scatter in units of area. In the context that RCS is commonly applied, the detection or measurement of metallic vehicles, the RCS is approximately equal to the equivalent surface of reflection, since perfectly electric conductors will re-scatter radio waves as single, coherent surface. 

In our scenario, a hypothetical radio reflection of a cloud, a large volume of very low density of water molecules, there is no resemblance between the RCS the physical area or dimensions in the cloud. However, regardless of the units of area, the radar cross section remains a measure of the power of reflection of any target, which can be interpreted as the equivalent efficiency of the radio reflection process. 
\begin{figure}
    \centering
    \includegraphics[width=1\linewidth]{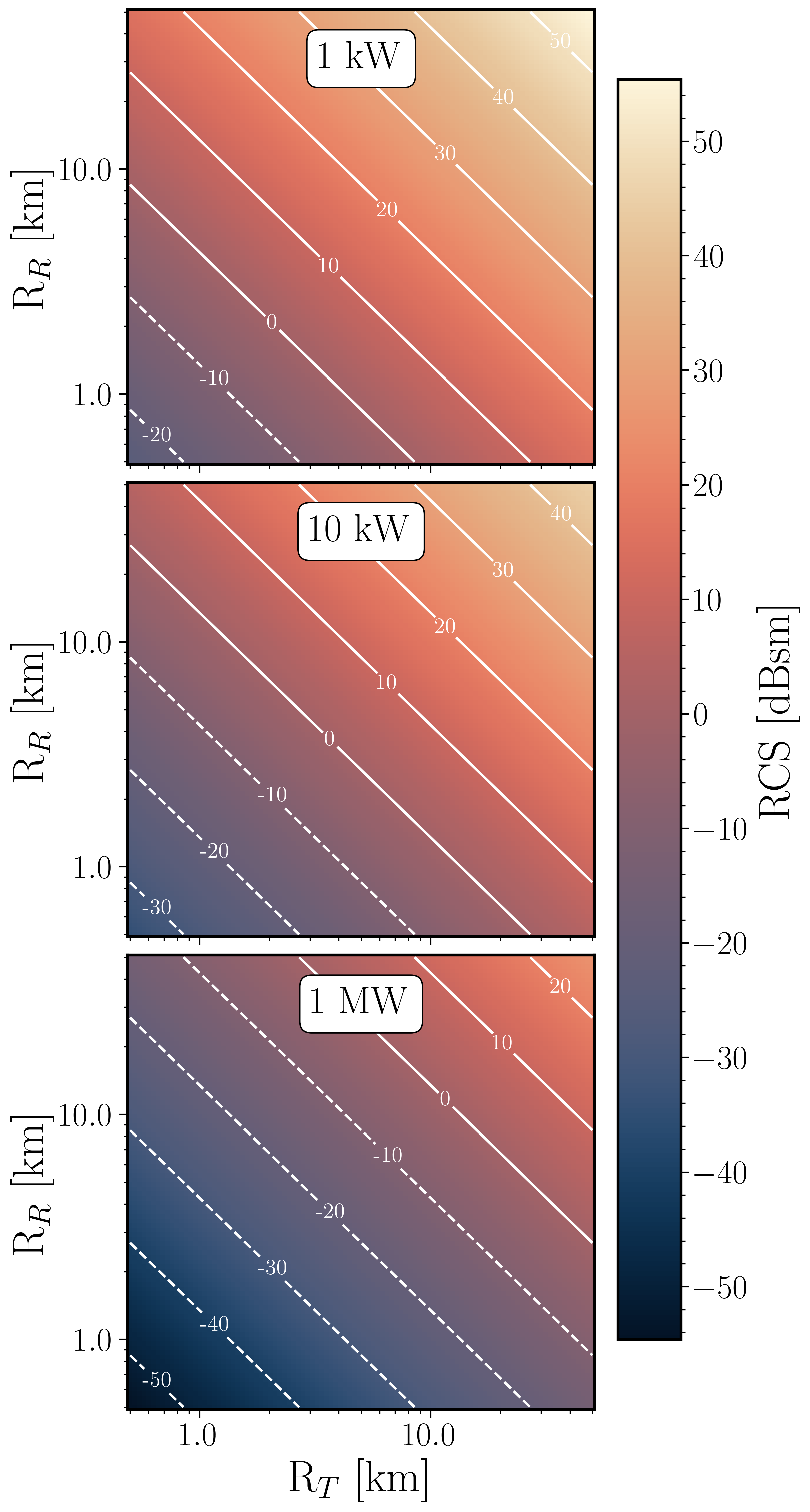}
    \caption{Radar cross section as function of receiver and transmitter distances $R_R$ and $R_T$ for different transmitter powers (1 kW (top), 10 kW (middle) and 1 MW (bottom)). All figures assume a detectable power of $P_R = -90 $ dBm at 50 MHz.}  
    \label{fig:RCS_vs_RT_RR}
\end{figure}

Fig.~\ref{fig:RCS_vs_RT_RR} provides an overview of different combinations of distances, RCS and power of transmitters. It shows that, for a powerful source, and a sufficiently small range, a detectable signal is realistic even under very small RCS, even down to 1 cm$^2 = -40$ dBsm. 

While one should assume that the RCS of a cloud was a long established fact, the issue, for this work, is the lack of available data or modelling in the MHz band. There is a great amount of meteorology science performed with radar techniques to monitor the atmosphere and the cloud cover. However, these studies are almost exclusively performed in the GHz frequencies. The GHz band offers a more powerful scatter signal ($\sigma \propto f^4$) and overall better resolution and focusing, plus the lack of strong backgrounds (like galactic and anthropogenic noise). This extends to the millimetre-wave region ($30-300$ GHz), where absorption lines of the water (vapour and liquid) are easily understood. 

There are radar instruments that study the atmosphere in the MHz band, for example, the incoherent scatter radars that map the ionosphere \cite{Woodman2019, Christakis2009}; or \textit{wind profilers}, which measure wind speed as a function of height \cite{Lehmann2021}. While some of these effects could be interesting to consider as a separate background, they provide no quantitative information useful for this work. Since there is, to our knowledge, no available model for radio reflection off a cloud in the MHz range, an existing radar model has been adapted for this use case. We do note that simulating a full cloud system remains a challenge even in cutting edge weather forecast.

%% file: sections/04_cloud_model.tex
In the following we present a simplified calculation of the RCS of a cloud. We note that this contains coarse assumptions simply to establish the feasibility of the reflection mechanism. The calculations are based on an extension of \cite{RadarEchoTelescope:2023oxh} and further details can be found in \cite{Cantarini2025}.

\subsection{Water vs Ice clouds}

We ought to start with a few important fundamental assumptions about the behaviour of hydrometeors as a basis of our modelling efforts. Hydrometeors span water droplets, free-falling raindrops, ice crystal and a mixture of all of the above. We only consider here liquid hydrometeors and treat them as perfectly spherical masses of water. This is because, at radio frequencies (MHz-GHz range, $f \in [3~\mathrm{MHz}~-3~\mathrm{GHz}], \lambda \in [0.1 - 100]~\mathrm{m}$), they fulfil the \textit{Rayleigh criteria}, where the size of the individual scatterer ($a$) in the cloud is (very) small compared to a wavelength ($a \ll \lambda$) \cite{Kleinman1965}. In this regime, the specific shape of the hydrometeor or its speed (relative to the timescale of the scattering radio wave) does not play a role, and it can be treated as a single coherent reflector. 
However, ice has intrinsic different EM properties than water; their relative permittivities are $\epsilon_{r,ice} \approx 3$,  $\epsilon_{r,water} \approx 80$~\cite{FujitaEtAl2000}. This makes for very different \textit{dielectric factors $|K|^2$} (from the Clausius-Mossoti relation) \cite{Smith1984}, where
\begin{equation}
    K = \frac{\epsilon_r-1}{\epsilon_r + 2}.
\end{equation}
Thus, $|K_{ice}|^2 = 0.17$ and $|K_{water}|^2 = 0.93$. A perfect electric conductor would have $|K_{pec}|^2 = 1$, making the water droplet a non-magnetic, dielectric medium much closer to a perfect reflector than the ice crystal. The low-density ice crystals or snowflakes are mostly composed of air can be treated as a weak perturbation on the air/vacuum, a ``soft" particle ($a~|\epsilon_{r,ice}-1| \ll 1$). In this case, a more applicable solution would be the Rayleigh-Gans approximation \cite{LeinonenEtAl2018}.

Therefore, the work presented in the following assumes liquid water drops and droplets acting as coherent, perfect dielectric radar scatter. The simulation of the radar scatter off of a droplet cloud was done using the Macroscopic Approach to the Radar Echo scatter (MARES) simulation package \cite{RadarEchoTelescope:2023oxh}. MARES is semi-analytic , first-principles radar scatter simulation software originally developed for the radar scatter off a relativistic ionisation trail in dense media. However, the basis of MARES lies on the general principle behind much of the modelling of radar scatter: The superposition of the linear electric fields scattered by Rayleigh regions. This means that, as long as the target can be decomposed as a collection of Rayleigh regions, MARES can be used to compute the total radar scatter. The calculation of a radar scatter of a complex object (such as a cloud) gets reduced to the problem of decomposing it into a series of Rayleigh regions.

\begin{figure*}
    \centering
    \includegraphics[width=1\linewidth]{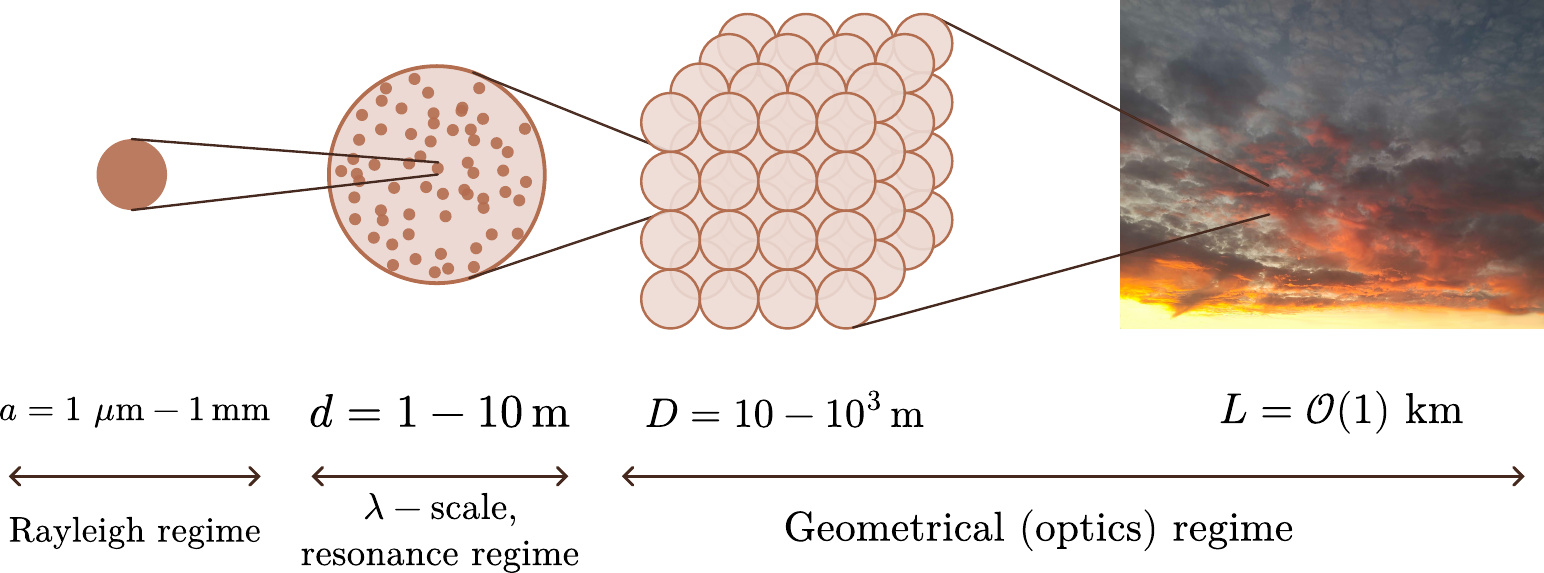}
    \caption{Schematic view of the challenge of cloud modelling. The relevant lengthscales span about $10^{18}$ orders of magnitude. Here, the goal is to model the cloud in as collections of arbitrary middle-scale unit volumes (cells). The downside is that there is no analytical solution for these cells, which are of size comparable to $\lambda$.}
    \label{fig:cloud_diagram}
\end{figure*}

\subsection{Radio scattering from a single droplet}

 The textbook solution (correct, yet naive) approach would be to attempt to simulate the radar scatter off of every individual water droplet. We know that, for a single droplet, the relation between the incident and scattered fields is 
\begin{equation}
     \begin{pmatrix}
E_{s\parallel} \\
E_{s\perp}
\end{pmatrix}
=
\begin{pmatrix}
S_2 & S_3 \\
S_4 & S_1
\end{pmatrix}
\frac{e^{-ik(r-z)}}{ikr}
\begin{pmatrix}
E_{i\parallel} \\
E_{i\perp}
\end{pmatrix}
 \end{equation}
with the incident and scattered fields, $E_i$ and $E_s$, decomposed in perpendicular y and parallel terms to a reference frame and $S$ being the amplitude scattering matrix. Under spherical approximation, $S_3 = S_4 = 0$ and the relationship between the amplitude scattering matrix $S$ and the radar cross section is:
\begin{equation}
\label{eq:RCS_as_function_of_scattering_amplitude}
    \sigma = \frac{\lambda^2}{\pi}|S|^2 =\frac{\lambda^2}{\pi}\frac{|S_2E_{i,||}|^2 + |S_1E_{i,\perp}|^2}{|E_{i,||}|^2 + |E_{i,\perp}|^2}.
\end{equation}
This simply means that $S_1$ and $S_2$ respectively describe perpendicular and parallel polarisation: $\sigma_{\perp, \parallel} = 4\pi|S_{1,2}|^2/k^2$. Under Rayleigh conditions, we can describe the amplitude coefficients in terms on the polarisability of the sphere $\alpha$; $S_1 = 2i\pi^2\alpha/\lambda^3\epsilon_0$ and $S_2 = 2i\pi^2\alpha\cos(\theta)/\lambda^3\epsilon_0$. In turn, $\alpha=4\pi\epsilon_0a^3K$ depends on the sphere radius $a$, the absolute permitivity of free space $\epsilon_0$ and the dielectric factor $K = 0.93$, as above. In the textbook case, all that is required is to combine the radar cross section of the droplet with Eq.~\ref{eq:radar_range}. The total amount of instantaneous power scattered at the antenna is the sum of all illuminated droplets for that instant. 

The most simply type of cloud we can consider is a \textit{monodisperse} cloud, a cloud comprised of droplets that have the same size and mass. A cloud is defined by its \textit{liquid water content} (LWC), where, in the monodisperse case, is proportional to the number density of droplets $n$:
\begin{equation}
    \mathrm{LWC} = \frac{M_\mathrm{cloud}}{V_\mathrm{cloud}} = \frac{N}{V} \cdot m_\mathrm{droplet} = n\,
    \left(\frac{4}{3}\pi a^{3}\right)\rho_\mathrm{water},
\end{equation}
with $\rho_\mathrm{water} = 10^{6}\ \mathrm{g\,m^{-3}}$. A monodisperse cloud has also a simply-defined bulk radar reflectivity factor, $Z = n(2a)^6$, which can be measured with weather radars. 

In this work, we use 3 different examples of monodisperse clouds (shown in Table~\ref{tab:cloud_droplets}), with typical values derived from the literature~\cite{MarshallPalmer1948, Miles2000, Gultepe2007}. 

\begin{table}[htbp]
  \centering
  \begin{tabular}{lcccc}
    \toprule
    Label    & $a$        & $n$ [m$^{-3}$]  & LWC [g\,m$^{-3}$] & $Z$ [dBZ] \\
    \midrule
    Rain    & $0.5$\,mm  & $1\times10^{3}$ & $0.52$ & $30$  \\
    Stratus & $10\,\mu$m & $1\times10^{8}$ & $0.42$ & $-22$ \\
    Fog     & $5\,\mu$m  & $2\times10^{8}$ & $0.10$ & $-37$ \\
    \bottomrule
  \end{tabular}
    \caption{Droplet parameters chosen for our simulations.}
  \label{tab:cloud_droplets}
\end{table}
``Stratus" stands for stratocumulous, low-altitude, non-precipitating clouds. They show the highest reported variation in 
$a$ and LWC. The values chosen here are above the average reported in \cite{Miles2000}, but still realistic. Monodisperse clouds are far from reality, but, in practice, the scatter will be dominated by the strongest component, as evidenced by the estimated $Z$ values. 

It is estimated that the average droplet density in a cloud is of the order of $n = [5\times10^{7} - 3\times10^{9}]~\mathrm{m}^{-3} = [50 - 300]$ droplets~$/\mathrm{cm}^3$~\cite{GryspeerdtEtAl2022}. Fully simulating the radar scatter off a volume of 1 $\mathrm{km}^3$ (only a small cloud) would require calculating $O(10^{17}-10^{18})$ individual droplets. The means that calculating individual sources is not a method scalable to large clouds. It is worth noting that the problem of length-scales in simulation of cloud physics is one of the large and long-standing open problems in the field e.g.~\cite{MorrisonEtAl2020a}, and is a topic of active research, with a breadth of different simulation approaches to predict the effects of the micro-scale droplet formation on large-scale clouds. In order to get a first estimate of the problem of MHz scattering off of clouds from scratch, we need to make it conceptually easy and computationally feasible. 

Relaxing the scales of the problem can reduce the number of operations: at our wavelengths of interest (0.1 m - 10 m), the Rayleigh region requisite ($d \ll \lambda$) can reasonably be assumed be as large as 1 cm, larger than a single droplet, but insufficient to eliminate the problem of large-scale simulation ($O(10^{15})$ operations). In our simulation method (see Figure \ref{fig:cloud_diagram}) the $10^{18}$ orders of magnitude between the droplet and the large cloud are split in two steps, around the $1 - 10~\mathrm{m}$ length-scale. The full cloud is divided in these \textit{unit volume} sections (or cells), and where all cells are identical volumes of randomly distributed water droplets. The only difference between the scatter of the different cells is due to their relative position and geometry, which can be computed with MARES. 
It should  be noted that taking scattering from regions larger than $d\sim 10$ cm means that they no longer fulfil Rayleigh conditions. 

\subsection{Modelling a cloud cell}

\begin{figure}
    \centering
    \includegraphics[width=1\linewidth]{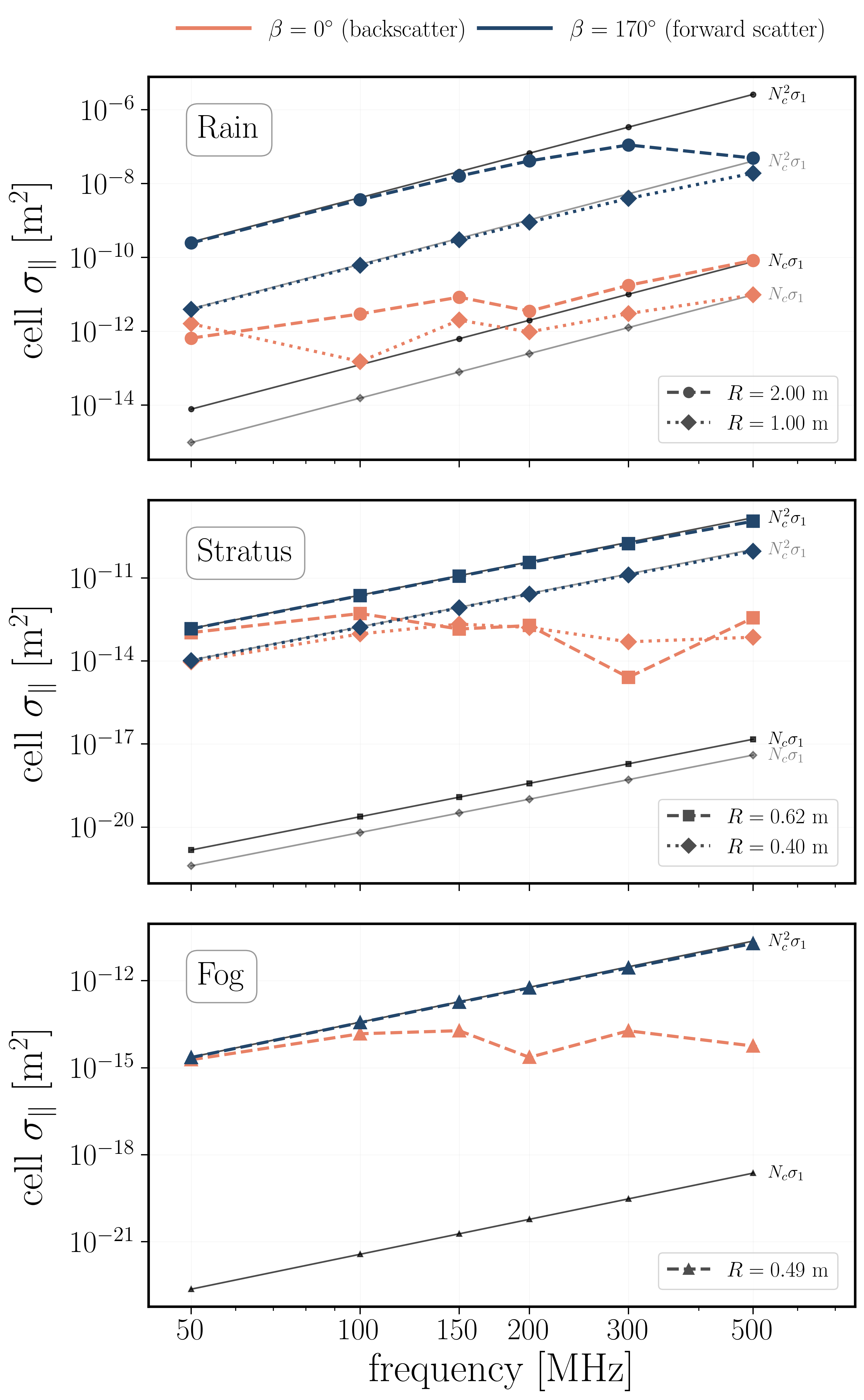}
    \caption{The characteristic parallel RCS ($\sigma_\parallel$) of the different cloud cell configurations at different frequencies and orientations. The TX-cell-RX plane is called the scatter plane, and determines the chosen ($\sigma_\parallel$), with the electric field polarisation in the plane, and the \textit{bistatic angle} $\beta$ as the aperture between TX and RX as seen from the sphere's centre. 
    Thus, $\beta= 0$ is the perfect backscatter, the scenario where TX and RX are co-located. $\beta=170^{\circ}$ is the near-forward reflection, our geometry of interest. For an equidistant cell from TX and RX, $\beta=170^{\circ}$ means that the cell sits $5^\circ$ off the line of sight between TX and RX (low elevation). The full lines show the fully incoherent ($N_c\sigma_1$) and the fully coherent ($N_c^2\sigma_1$) scattering limits.}
    \label{fig:cell_models}
\end{figure}

The first step in the full cloud simulation is to model the radio scattering of a medium-sized cloud cell from individual droplets. As the cloud cells are an arbitrary choice imposed by us, we can freely take them as spheres, in order to exploit their symmetry and obtain two independent polarisation modes (as shown in Eq.~\ref{eq:RCS_as_function_of_scattering_amplitude}). Since the target size of the cloud volume is $O(0.1 -1)$ m, comparable to the radio wavelength $\lambda$, we are in the so-called \textit{Mie} regime. Therefore, the RCS of the cloud cells was chosen to be determined numerically via MARES simulations of droplets.

\begin{table}[t]
\centering
\begin{tabular}{lcccccc}
\toprule
Cell & $a$ & $d$ [m] & $V$ [m$^3$] & $N_c$ & trials \\
\midrule
Rain    & 0.5\,mm      & 2.0   & 33.5  & $3.4\times10^{4}$ & 10--400 \\
Rain    & 0.5\,mm      & 1.0   & 4.19  & $4.2\times10^{3}$ & 2--400 \\
Stratus & 10\,$\mu$m   & 0.620 & 1.00  & $1.0\times10^{8}$ & 2--4 \\
Stratus & 10\,$\mu$m   & 0.400 & 0.268 & $2.7\times10^{7}$ & 2 \\
Fog     & 5\,$\mu$m    & 0.492 & 0.50  & $1.0\times10^{8}$ & 2--4 \\
\bottomrule
\end{tabular}
\caption{Values of cells used to compose the clouds. Each calculated cell is a sphere of radius $d$ and volume $V=\tfrac{4}{3}\pi d^3$ holding $N_c = nV$ droplets of radius $a$. For the low-droplet count cases, the average
over several trials is taken to reduce artifacts; the number of trials varies with frequency.}
  \label{tab:cloud_cells}   
\end{table}

We ran a number of different MARES simulations of groups of droplets arranged in different spherical volumes at different radii between $d = 0.5$ and $d = 2$ m. The different cell parameters have been summarised in Table~\ref{tab:cloud_cells}. Each simulation, a full transmitter-cell-receiver (TX-cell-RX) combination, is computed for 50 equidistant RX positions in the scattering plane ($\Delta \theta = 7.2^\circ$), and, for each position, the two antenna orientations, in plane ($\sigma_{\parallel}$) and out of scattering plane ($\sigma_{\perp}$). A total distance of 600 m between TX and RX and the sphere ensures that the computed RCS includes far-field effects at these frequencies only. The chosen sphere radius is chosen to keep the density consistent with the definition in Table~\ref{tab:cloud_droplets} while effectively being constrained by the memory requirements of the MARES calculation. It was determined that a limit of $N_c \le 10^{8}$ kept the simulations practical on the available computing infrastructure, at the cost of limited sphere size.

These cell simulations were run at different operating frequencies $f = [50, 100, 150, 200, 300, 500]~\mathrm{MHz}$. An sample of the results is presented in Fig.~\ref{fig:cell_models}. Rain and stratus models were ran with two different radii to test for consistency. The most basic expectation from theory, the RCS dependence on the droplet radii, is evident where with the rain simulations having a much stronger response than stratus configuration and stratus a stronger one that the fog. It is quite apparent that fog-like water configurations give a very weak radar scatter. We included the fog here for comparison and conceptual understanding.

Additionally, we know that every sphere must present an RCS constrained between fully incoherent droplet sum, where $\sigma_C \propto N_c \sigma_a$, and a fully coherent droplet sum, where $\sigma_C \propto N_c^2\sigma_a$. We include those lines as a visual guide in Fig.~\ref{fig:cell_models}, and they outline two interesting behaviours. First, the backscatter geometries (where the RX is co-located with the TX) show 

Despite being at the same total distance from the cell, the contributions of the individual, randomly placed droplets within the sphere largely cancel each other and the RCS, effective scatter strength, is much smaller. We also notice non-monotonic behaviour in the backscatter curves, like the dip around 200 MHz the fog configuration, which are expected behaviour for a non-solid sphere in a Mie-like regime. The round trip path difference across the distribution matches very closely the wave-vector at these frequencies, so the scattering is suppressed.

The most significative conclusion is that at forward scatter geometries, the sphere cell is shown to be strongly coherent, overlapping with the coherent limit at most configurations and frequencies. Only rain in the sparsest cell configuration and at the smallest wavelengths is ``seen" as incoherent under forward scatter. 

\subsection{The radar cross section from a cloud}
\begin{figure}
    \centering
    \includegraphics[width=1\linewidth]{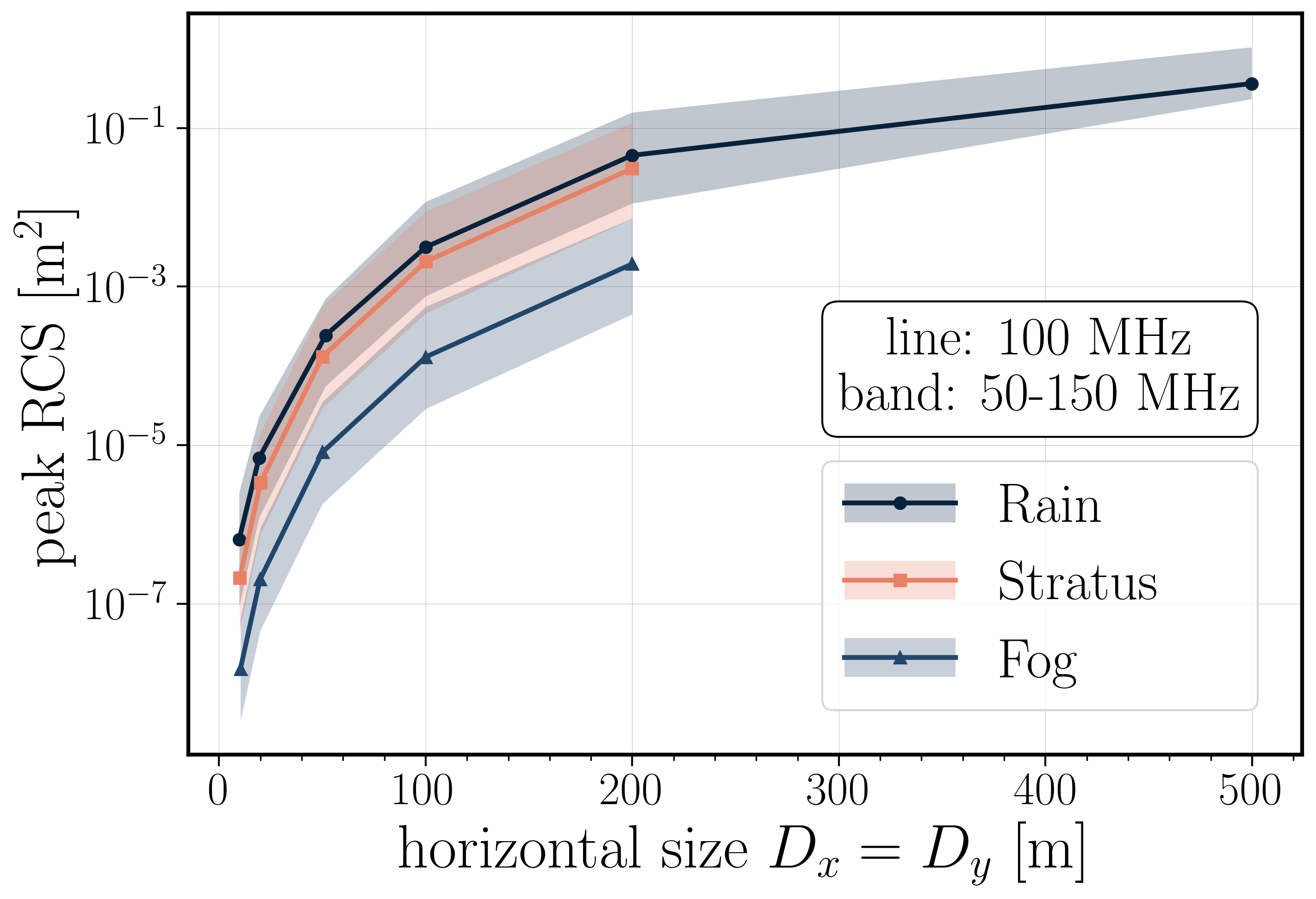}
    \caption{Maximum value of the radar cross section as a function of lateral cloud size, as computed by MARES. The lines show the results for the three cloud types simulated at a frequency of $f=100$ MHz, with the band showing the range of values between the simulations at 50 MHz (lower bound) and 150 MHz (upper bound). These lines are the same as the ones presented in Fig.~\ref{fig:cloud_elevations}, in the fixed elevation case of $\epsilon = 5^\circ \equiv \beta = 170^\circ$ (forward scatter).  }
    \label{fig:cloud_frequencies}
\end{figure}
After obtaining the tabulated cell response for different configurations, we can build a full cloud volume as a group of cells, as presented schematically in Fig~\ref{fig:cloud_diagram}. Each simulated cloud is a cube grid of total horizontal size $D_x=D_y$ and depth $D_z$, rounded to whole cube. For the purposes of full cloud coverage, each cube is a cell of equal volume as the original sphere, with side $L_c = V^{1/3} = (4\pi/3)^{1/3}d \simeq 1.6d$. As a geometrical modelling choice, this leaves a region of $0.2~d$ of each sphere overlapping with each neighbour in all sides, and under-covered regions around the edges of the cube. This attempts to reflect the non-constant density distribution of the water distribution in the cloud. The actual spatial variability of droplets in within a cloud is which is seemingly very dependant on specific cloud conditions, making it another long-running research problem in the field, for which multiple modelling and experimental approaches exist \cite{Warner1979, Gerber2001, Wood2005, Dodson2022}.

Within MARES, each cell is treated as a point-like scattering centre with equivalent coordinates to its position, and the effects of the cell shape are encapsulated in the pre-computed RCS. Therefore, we also need to impose a jitter on the cell position over its geometrical centre. The jitter is taken as a Gaussian-distributed offset of standard deviation $\delta = L_c/3$ per axis, truncated at a maximum of $\pm L_c/2$. The jitter is meant to suppress the scattering enhancement effect that can happen due to the periodic and uniform positioning, an effect known as Bragg scattering. In practice, jitter also makes the real cloud distribution of droplets less uniform throughout the cloud. At a sufficiently high number of cells and jitter, any placement should be comparable to that of a random positioning within the cloud, while avoiding accidentally lower effective cloud sizes.

Each of the three types of simulated cloud cells presented above were used for the full cloud simulations. The characteristics for the cloud simulations are summarised in Table~\ref{tab:cloud_sweeps}. In this case, the rain configuration had the largest simulated cell volumes, which means that it makes for the least demanding cloud simulations, with 10 times fewer cells for a 200 m x 200 m x 100 m cloud volume than the other two cases. This meant that simulated rain clouds could easily be constructed up to larger sizes (500 m side) than the other two configurations. This is, once again, in an attempt to run a MARES simulation within reasonable time and memory constraints. 

\begin{table}[htbp]
  \centering
\begin{tabular}{lcccccc}
    \toprule
            & $L_c$ [m] & $\delta$ [m] & $D_{x,y}$ [m] & $N_{x,y}$ & $N_z$ & $N_{\mathrm{cells},\,200\,\mathrm{m}}$ \\
    \midrule
    Rain    & $3.224$ & $1.075$ & $10$--$500$ & $3$--$155$   & $31$  & $1.2\times10^{5}$ \\
    Stratus & $1.000$ & $0.333$ & $10$--$200$ & $10$--$200$  & $100$ & $4.0\times10^{6}$ \\
    Fog     & $0.794$ & $0.265$ & $10$--$200$ & $13$--$252$ &  $126$ & $8.0\times10^{6}$ \\
    \bottomrule
  \end{tabular}
  \caption{Simulation parameters for ``full'' clouds. 
  Each cube has a side size of $L_c$ for an equivalent volume of the cell calculated in Table~\ref{tab:cloud_cells}. $D_z = 100$\ m has been used for all cloud types.  
  $N_{x,y,z}$ is the number of cell layers in each dimension.
  The last column is the total number of cells at $D_{x,y}\approx200$\,m.}
  \label{tab:cloud_sweeps}
\end{table}
\begin{figure*}
    \centering
    \includegraphics[width=1\linewidth]{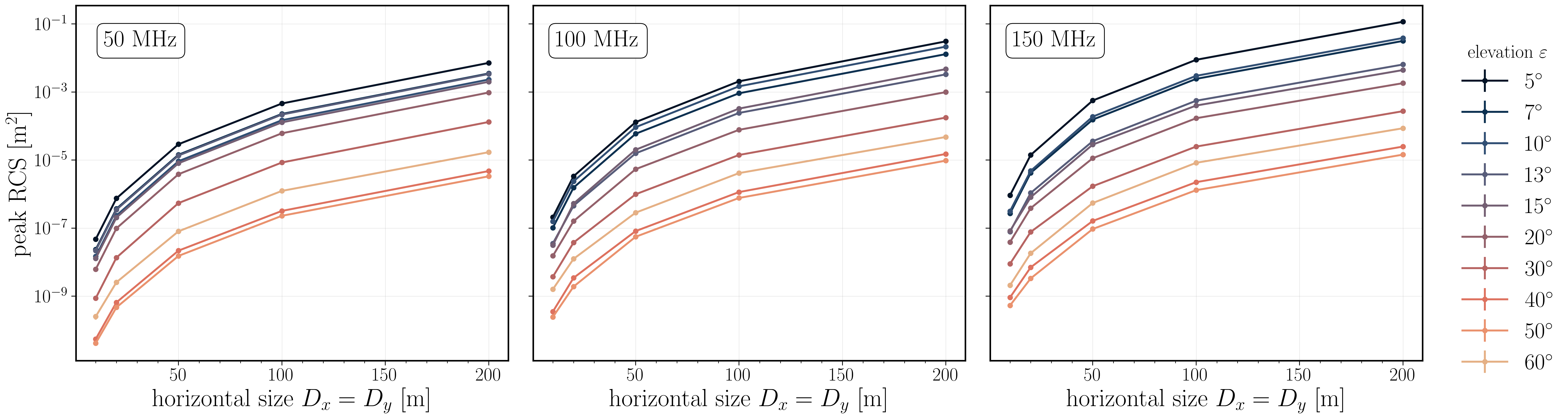}
    \caption{Maximum value of the radar cross section as a function of lateral cloud size, for a stratus cloud configuration and a fixed cloud thickness of $D_z = 100~\mathrm{m}$. The geometry is the same as for Fig.~\ref{fig:cloud_frequencies}.}
    \label{fig:cloud_elevations}
\end{figure*}
Fig.~\ref{fig:cloud_frequencies} shows a high-level result for the three cloud configurations simulated in this work; with Fig.~\ref{fig:cloud_frequencies} providing a closer look into the geometrical dependency of the different simulations. Both figures show the peak RCS value of the MARES simulation for a fixed bistatic geometry and variable cloud lateral (horizontal) size, for three different simulation frequencies in the VHF band. The cloud was placed centred between the TX and RX, at equal horizontal distance to both and an elevation such that both antennas have the same viewing angle $\epsilon$ from their baseline.

For Fig.~\ref{fig:cloud_frequencies}, the results are chosen at a fixed elevation of $\epsilon = 5 ^\circ$, which makes the bistatic angle to be ($\theta = 180^\circ - 2\epsilon = 170^\circ$), putting the centre of the cloud volume in the forward-scatter geometry. The main results of this figure is that we obtained peak (instantaneous) radar cross sections of approximately $0.1~\mathrm{m}^2 = -10~\mathrm{dBsm}$, which are compatible with our previous calculation of a detectable threshold signal, presented in Fig.~\ref{fig:RCS_vs_RT_RR}. Moreover, this result holds for both rain and stratus configurations, strongly hinting to the fact that this result holds regardless of modelling parameter choices like cell size or count, and that it depends on bulk cloud properties like the LWC. 

In Fig.~\ref{fig:cloud_elevations}, the stratus cloud configuration is plotted using the same axes as Fig.~\ref{fig:cloud_frequencies}, but separating explicitly into different frequencies and apparent cloud elevations $\epsilon$. We notice two important results: First, the RCS increases with increasing frequency (decreasing wavelength), as expected from a scattering model based on Rayleigh scattering, and second, we notice a very strong enhancement (multiple orders of magnitude) of the radar cross section at low elevations. This can be understood as a radio equivalent to the known Fresnel grazing effect, where reflectivity is strongly enhanced at near parallel angles to the surface. For this to be true, our collection of cells modelled by MARES must have have been sufficiently to calculate, from individual points, a bulk effect from a large cloud surface.

%% file: sections/05_conclusions.tex
We find that there is currently no satisfactory explanation for the near-horizon event clusters that have been independently reported by self-triggered radio experiments, such as AERA, OVRO-LWA, BEACON, and GRAND. The aim of this work is to highlight their existence, and attempt to understand their nature. We conclude that these horizon events are conceptually compatible with radio reflections of accidental transmitters on nearby clouds. We presented estimates of the strength of the transmitter and the relevant distances using available data, albeit limited, and reasonable assumptions about the sensitivity of the instruments. These estimates depend on the strength of the radar scatter off clouds in the MHz bands, which are usually ignored in the meteorological literature in favour of the GHz bands. 

MARES \cite{RadarEchoTelescope:2023oxh}, a general-purpose model developed for the radio astroparticle community, has been adapted to estimate the radar cross-section of clouds in the MHz regime. A cloud model has been built from individual cloud droplets as dielectric scatterers, first constructing droplet-filled cells and then assembling large-scale clouds. This model serves for a first-order approximation of a cloud, beginning with a single size of droplet per cloud and making multiple compositions of simple elements, which clearly does not reflect the complexity of real clouds. Therefore, the results presented here must not be taken as anything other than a first attempt. 

However, using it as a general feasibility assessment of radio reflection properties of clouds, we have found that the peak radar cross section lies between $10^{-10}-10^{-3}\ \mathrm{m}^2$ and is, thus, compatible with a detection of a reflected transmitter in the near-horizon geometry. In fact, we have found an enhancement of the reflectivity of a large-scale cloud volume, a property of the cloud bulk, from the interference of contributions of many cloud cells. This enhancement happens at exactly the forward scatter geometries that are consistent with a near-horizon signal. In this work, we thus cannot confidently reject the hypothesis that low-elevation cloud reflections can act as background for MHz radio detectors. Future work, experimentally and in modelling is necessary to confirm this potential effect.

We encourage experiments to be more thorough in their study of backgrounds, considering the possibility that atmospheric phenomena influence radio backgrounds. For experiments needing a $\mathcal{O}(10^{-9})$ suppression of backgrounds, physical origins need to be considered to be effectively mitigated. 
We note that our approach focusses on visible clouds. In researching properties of such clouds, we encountered a number of other atmospheric phenomena that could be relevant in the MHz regime, such as clear-air turbulence and winds e.g.~\cite{Ralph1995, McDonald2006,Kumar2019,Lehmann2021, Wu2026}. These should be considered experimentally and included in further modelling work. Obviously, further modelling work also needs to include full wave-forms, multi-path calculations, and triggering details. 

For the sake of efficiently working self-triggers, we hope that our anecdotal and model-driven hypothesis can be rejected experimentally. 

%% file: sections/A_radio_propagation_atmosphere.tex
The atmosphere is a medium with a slowly-varying index of refraction, and, as such, is capable of bending the light and radio waves that propagate through it. This effect is more present in the lower atmosphere ($< 5$~km), where the gradient is higher. A thin, $O(10-100)$~m, gradient layer that is capable of bending low-frequency radio waves, changing their direction and extending their range is commonly called a duct. A simple estimate of the typical deviation expected from an atmospheric duct is presented here. Under the ray optics framework, the total deviation from the starting direction $\theta$ is just the integral of the curvature $\kappa$ along the path of the ray,
\begin{equation}
    \theta = \int_0 ^R \kappa ds.
\end{equation}
The curvature, which comes from the eikonal equation, relates the change of the direction of the ray to the gradient of index of refraction in its perpendicular component ($\vec{N}$):
\begin{equation}
 \kappa = \frac{1}{n} \nabla n \cdot \vec{N}
 \rightarrow
 \kappa(z) = \frac{1}{n(z)} \frac{dn}{dz}.
\end{equation}
Which, in this case, it can be simplified to one dimension since $n = n(z)$ and the ray is assumed moving horizontally for its path ($\vec{N} = \hat{z}$). This implies that ray deviations are small, in other words, the curvature does not change significantly along the path: $\kappa \neq \kappa(s)$ and it only depends on the local gradient $\nabla n$. The consequence is that, for air ($n \simeq 1$), we have

\begin{equation}
    \theta \cong \kappa R = \frac{1}{n(z)} \frac{dn}{dz} R \rightarrow \theta \approx \frac{dn}{dz} R.
\end{equation}

This approximation, albeit coarse, should hold for typical duct layers over long horizontal distances. In order to calculate the index of refraction gradient in the atmosphere $dn/dz$, we follow the official ITU recommended parametrization \cite{ITU2019}, where $N = (n - 1) \cdot 10^6$, 
\begin{align}
    N(z) &= 315 * e^{\frac{-z}{7.5~\text{[km]}}}, \\
    \frac{dN}{dz} &= -\frac{1}{7.5} N(z) \cong -42 * e^{\frac{-z}{7.5~\text{[km]}}} ~~ \left[\frac{\text{N-units}}{\text{km}}\right].
\end{align}
 The exponential term $e^{\frac{-z}{7.5~\text{[km]}}} < 1$ is evaluated term is the height of the duct layer a.s.l. As we already assumed above that the values is slowly-changing, we also can approximate it as constant. That makes the deviation angle
 \begin{equation}
    \theta 
    \approx \frac{dn}{dz} R
    \approx \frac{dN}{dz} R \times 10^{-6}
    \simeq - 4 R([\text{km}]) \times 10^{-5}.
\end{equation}

Thus, a typical deviation over a range of $R = 100$~km is $\theta \sim 0.2^\circ$. A conservative value of the maximum deviation would be 10x, or $\theta_{max} \sim 2$ degrees. This corresponds to 10x stronger local gradients ($dN/dz \sim 10^3$) over a short range, accounting for e.g.\ extreme temperature layers. However, strong gradients are incompatible with this framework as they would anyway quickly deviate the ray out of the duct, making the approximations used here invalid.

A more detailed calculation, using Smith-Weintraub's updated formula \cite{Rueger2002}, expresses $N$ as a function of pressure and water content. This was calculated in, e.g., \cite{Kanitz2020}, but the conclusions do not change that atmospheric ducting is insufficient to explain the excess. 